\documentclass[sigconf]{acmart}

\setcopyright{none}

\newcommand{\quotebox}[2]{\textit{``#2''}\ifthenelse{\equal{#1}{}}{}{ \mbox{-}~#1}}

\usepackage{color}
\usepackage{nameref}
\usepackage{multirow}
\usepackage{tabularx} 
\usepackage{censor}
\usepackage{nameref}

\usepackage{mdframed}

\usepackage[most]{tcolorbox}

\usepackage{xcolor}          

\definecolor{lightblue}{RGB}{0, 0, 100}

\newtcolorbox{MyBox}{
  colback=white,
  colframe=lightblue,
  fonttitle=\bfseries,
  coltitle=black,
  sharp corners,
  boxrule=1pt,
  left=5pt,
  right=5pt,
  top=5pt,
  bottom=5pt,
  breakable
}

\definecolor{purplish}{HTML}{D8DFE3}
\definecolor{purplishlight}{HTML}{EBEFF3}
\definecolor{purplishdark}{HTML}{FF7F50}
\definecolor{purplishdark}{HTML}{FF7F50}
\definecolor{myorange}{HTML}{FF7F50}

\newtcolorbox[auto counter,number within=section]{rqbox}[2]{
    nameref=#1,
    title=\small{#1}, 
    enhanced,
    attach boxed title to top left={yshift=-6pt, xshift=8pt},
    boxed title style={size=small,boxsep=1pt},
    colframe=purplishdark,colback=white,colbacktitle=purplishdark,
    boxsep=2pt,left=2pt,right=2pt,top=6pt,bottom=2pt,middle=2pt
}

\newcommand{\rqtextone}{
How are software professionals using LLMs to support coding activities, and what are their perceptions of these tools in practice?
}

\newtcolorbox[auto counter,number within=section]{promptbox}[2]{
    nameref=#1,
    title=\small{#1}, 
    enhanced,
    attach boxed title to top left={yshift=-6pt, xshift=8pt},
    boxed title style={size=small,boxsep=1pt},
    colframe=myorange,             
    colback=white,                 
    colbacktitle=myorange,        
    boxsep=2pt,left=2pt,right=2pt,top=6pt,bottom=2pt,middle=2pt
}

\definecolor{lightpurple}{RGB}{255, 229, 204}
\definecolor{darkpurple}{RGB}{255,140,0}

\newmdenv[
  backgroundcolor=lightpurple,
  linecolor=darkpurple,
  linewidth=3pt,
  leftline=true,
  topline=false,
  bottomline=false,
  rightline=false,
  innerleftmargin=10pt,
  innerrightmargin=10pt,
  innertopmargin=5pt,
  innerbottommargin=5pt,
  skipabove=10pt,
  skipbelow=10pt
]{participantbox}

\begin{document}

\title{Model-Assisted and Human-Guided: Perceptions and Practices of Software Professionals Using LLMs for Coding}

\author{Italo Santos}
\affiliation{%
  \institution{University of Hawai‘i at Mānoa}
  \city{Honolulu}
  \state{HI}
  \country{USA}
}
\email{italo.santos@hawaii.edu}

\author{Cleyton Magalhaes}
\affiliation{%
  \institution{UFRPE}
  \city{Recife}
  \state{PE}
  \country{Brazil}
}
\email{cleyton.vanut@ufrpe.br}

\author{Ronnie de Souza Santos}
\affiliation{%
  \institution{University of Calgary}
  \city{Calgary}
  \state{AB}
  \country{Canada}
}
\email{ronnie.desouzasantos@ucalgary.com}

\begin{abstract}
Large Language Models have quickly become a central component of modern software development workflows, and software practitioners are increasingly integrating LLMs into various stages of the software development lifecycle. Despite the growing presence of LLMs, there is still limited understanding of how these tools are actually used in practice and how professionals perceive their benefits and limitations. This paper presents preliminary findings from a global survey of 131 software practitioners. Our results reveal how LLMs are utilized for various coding-specific tasks. Software professionals report benefits such as increased productivity, reduced cognitive load, and faster learning, but also raise concerns about LLMs' inaccurate outputs, limited context awareness, and associated ethical risks. Most developers treat LLMs as assistive tools rather than standalone solutions, reflecting a cautious yet practical approach to their integration. Our findings provide an early, practitioner-focused perspective on LLM adoption, highlighting key considerations for future research and responsible use in software engineering.
\end{abstract}



\maketitle

\section{Introduction}

Large Language Models (LLMs) have rapidly gained ground in software engineering practice, becoming a prominent part of modern development workflows~\cite{ozkaya2023application, fan2023large, hou2024large, zheng2025towards}. With the release of models such as Codex and ChatGPT, practitioners have started to adopt LLMs to support various development tasks throughout the software lifecycle. Recent studies show that professionals are incorporating these tools into daily workflows, particularly for assistance and experimentation~\cite{khojah2024beyond, nam2024using}. Previous studies have demonstrated the use of LLMs across major areas of software development, including requirements engineering, software design, development, quality assurance, maintenance, and project management~\cite{hou2024large, zheng2025towards}. Within these domains, LLMs have been applied to tasks such as specification generation, code summarization, program repair, test case generation, vulnerability detection, and documentation~\cite{wang2024software, hou2024large, zheng2025towards, santos2024we}. In addition to backend tasks, LLMs have also been integrated into development environments, providing in-IDE support for code understanding and information generation~\cite{nam2024using}. 

These applications demonstrate the versatility of LLMs as both automation engines and interactive assistants within the software lifecycle. Yet, coding activities appear to be the most common context in which LLMs are actually used by practitioners~\cite{khojah2024beyond, nam2024using}. Therefore, to better understand how these tools are being adopted and experienced in real-world settings, this study presents our preliminary findings on how practitioners utilize LLMs in coding activities, focusing on the experiences of software professionals from around the world and the types of support they seek, as well as their perceptions of these tools in practice. The aim of our study is to identify current usage patterns, reveal practical concerns, and inform future strategies to improve the integration of LLMs—tools that are undoubtedly becoming a core part of the software development lifecycle. To guide our research and its application in an industrial and practical context, we formulated the following question:

\newcommand{\rqone}[2][]{
     \begin{rqbox}{\textbf{Research Question}}{#2}
         \rqtextone
         #1
     \end{rqbox}
 }

 \rqone{}

To answer our research question, we conducted a survey with 131 software professionals, targeting individuals with real-world experience using LLMs. Our survey combined open and closed questions to capture detailed accounts of how LLMs are used in practice, including specific coding tasks, perceived benefits, limitations, and integration into daily development routines. We used thematic analysis to identify common usage patterns and perceptions of practitioners, supported by descriptive statistics to contextualize our findings. Our findings suggest that most developers use LLMs as assistive tools rather than final solutions: modifying generated code, seeking explanations, or brainstorming while coding manually, while only a minority directly copy-paste code. Although LLMs are valued for speeding up tasks and supporting learning, developers remain cautious due to their limitations, favoring human oversight. This paper makes three primary contributions:
\begin{itemize}
    \item We present a global survey that captures how software professionals use LLMs to support their coding activities.
    \item We identify key benefits and challenges reported by professionals when integrating these tools into their development workflows.
    \item We document practical experiences and unmet needs expressed by practitioners, providing a grounded understanding of how current LLM tools align with or fall short of real-world coding practices.
\end{itemize}

Based on this introduction, the study is organized as follows. Section~\ref{sec:method} describes our survey methodology. In Section~\ref{sec:findings}, we present our findings, which are discussed in Section~\ref{sec:discuss}. Section~\ref{sec:limitations} discussed threats to validity. Finally, Section~\ref{sec:conclusions} summarizes our contributions and final considerations.

\section{Background} 
\label{sec:background}

The study of LLMs in software engineering is rapidly evolving, as demonstrated by several recent literature reviews and surveys~\cite{fan2023large, hou2024large, michelutti2024systematic, wang2024software, zheng2025towards}. Early work in this area mainly focused on the potential of LLMs to support isolated development tasks. However, more recent reviews reveal a growing interest in broader applications, with LLMs now being considered at multiple stages of the software development life cycle and in diverse professional contexts~\cite{fan2023large, hou2024large, michelutti2024systematic, wang2024software, zheng2025towards}. This expanding body of work reflects an increasing interest in understanding not only the technical capabilities of LLMs but also their integration into real-world workflows~\cite{hou2024large, zheng2025towards, fan2023large}. As the research community continues to expand its scope, LLMs are increasingly being framed as general-purpose tools with the potential to influence multiple aspects of software engineering practice~\cite{fan2023large, hou2024large, wang2024software}.

Building on this foundation, recent empirical research has focused on how software professionals actually use LLMs in practice. Studies have explored how developers integrate these tools into their workflows, the types of support they seek, and their perceptions of usefulness and reliability~\cite{khojah2024beyond, nam2024using, rasnayaka2024empirical, santos2024we, jahic2024state}. Experimental work with IDE-integrated assistants further suggests that context-sensitive, prompt-less interactions can support code understanding, although perceived benefits vary depending on previous experience~\cite{nam2024using}. Focusing on the perspective of individuals, a recent global survey of professional testers revealed that even those without direct LLM experience recognized their potential to support tasks such as test design, documentation, and debugging \cite{santos2024we}. Practitioners reported using LLMs for test case generation, bug reproduction, and code clarification. Furthermore, respondents with limited programming backgrounds found LLMs helpful in bridging technical knowledge gaps, particularly during test analysis. However, concerns were raised about the lack of control over LLM outputs, risks of incorrect suggestions, and the absence of formal usage guidelines. 

These findings reflect both enthusiasm and caution in current practice, demonstrating the importance of aligning LLM capabilities with the specific needs of practitioners, at least in the context of software testing~\cite{santos2024we}. At the same time, they reveal a broader need to conduct more global, practitioner-centered studies that examine LLM usage in other specific areas of software development. As adoption increases, understanding how professionals interact with these tools across diverse software engineering activities will be essential to ensure that LLM-based solutions are both context-aware and grounded in real-world development practices.

\section{Method} 
\label{sec:method}

We investigated how software engineers use LLMs such as ChatGPT, Copilot, Bard, and Claude in real-world coding tasks. Following recent survey-based research in software engineering practice~\cite{hu2022practitioners, santos2024we, voria2025fairness}, we conducted a cross-sectional online survey~\cite{pfleeger2001principles, easterbrook2008selecting, ralph2020empirical} to explore the usage patterns, perceived advantages, and limitations of LLMs, and their integration into software development workflows, based on the experiences of software practitioners. In the following subsections, we describe the survey instrument, participant recruitment, and data analysis procedures.

\subsection{Survey Design}


We designed an anonymous survey composed of four thematic sections and one demographic section. The survey was implemented using Qualtrics~\footnote{\url{www.qualtrics.com}}, and all questions were formulated to capture real-world practices with LLM tools in the context of software development. The survey included both open-ended and closed-ended questions and was structured as follows:

\begin{itemize}
\item \textbf{Section I: LLM Usage in Coding Tasks.} This section asked participants to describe specific coding scenarios in which they used LLMs and to identify which LLMs and activities (e.g., debugging, implementation, testing) were involved.

\item \textbf{Section II: Workflow Integration and Benefits.} Participants were asked to describe how LLMs were integrated into their daily workflow and identify perceived advantages.

\item \textbf{Section III: Challenges and Limitations.} This section explored the drawbacks and potential concerns experienced by developers.

\item \textbf{Section IV: Demographic Information.} This final section gathered data on participants’ gender, country of work, years of experience, current role, and the types of software systems they work on.

\end{itemize}

The final survey questions were refined through internal reviews and structured to support both descriptive and thematic analysis.

\subsection{Pilot}
Before launching the survey, we tested the instrument on different devices and browsers to ensure functionality. We then conducted a small pilot with three software engineers working in industry. Their responses were not included in the final dataset. The pilot helped improve question clarity, confirmed relevance across professional roles, and ensured that the wording encouraged reflective and experience-based responses. Based on this feedback, minor adjustments were made to refine instructions and ensure a better fit for industry practitioners. Table~\ref{tab:questionnaire} shows the validated instrument considering the questions used in this preliminary study.

\begin{table}[ht]
  \caption{LLM Coding Survey Questionnaire}
  \label{tab:questionnaire}
  \begin{tabularx}{\linewidth}{p{1.8cm} X}
    \toprule
    Section & Questions \\
    \midrule

    Uses & 
    1. Describe a specific situation when you used an LLM (AI tool) to assist with a coding or programming task at work. \newline \\
    
    & 2. Which LLMs have you used for coding? (Select all that apply) \newline
    ( ) ChatGPT \quad ( ) GitHub Copilot \quad ( ) Google Bard / Gemini \quad ( ) Claude \quad ( ) Other (please specify) \newline \\
    
    & 3. Have you received formal training or education on AI/LLMs before using them for coding? \newline
    ( ) Yes \quad ( ) No \newline \\
    
    & 4. How do you integrate LLMs into your development workflow?  \\
    
    \midrule
    Advantages & 
    5. Describe any advantages you personally experienced when using LLMs for a coding task at work. \newline \\
    
    & 6. On a scale of 1 to 5, how effective do you find LLMs in improving your coding efficiency? \newline
    ( ) 1 \quad ( ) 2 \quad ( ) 3 \quad ( ) 4 \quad ( ) 5 \\
    
    \midrule
    Limitations & 
    7. Describe any challenges or limitations you personally experienced when using LLMs for a coding task at work. \newline \\
    
    & 8. What improvements would you like to see in LLMs for coding? \newline \\
    
    & 9. Would you be comfortable using LLM-generated code in production systems without human validation? \newline
    ( ) Yes \quad ( ) No \quad ( ) Only for non-critical tasks \\
    
    \midrule
    Demographics & 
    10. Select your gender: \newline
    ( ) Male \quad ( ) Female \quad ( ) Non-binary \quad ( ) Prefer not to say \newline \\
    
    & 11. Which country do you currently work in? \newline \\
    
    & 12. Years of experience as a software engineer: \newline
    ( ) Less than 1 year \quad ( ) 1–3 \quad ( ) 3–5 \quad ( ) 5–10 \quad ( ) More than 10 \newline \\
    
    & 13. What type of software do you work on? \newline \\
    
    & 14. What is your current role in the software industry? \\
    
    \bottomrule
  \end{tabularx}
\end{table}

\subsection{Data Collection}

We employed a crowdsourcing-based recruitment strategy to reach software professionals with experience using LLMs. The survey was distributed exclusively through the Prolific platform~\cite{russo2022recruiting}, a widely used tool in academic research that enables precise participant targeting. Pre-screening filters were applied to recruit participants who (1) reported working in software development roles, and (2) had experience using LLM tools such as ChatGPT. We selected the following Prolific filters: ``have knowledge of software development techniques'', ``have computer programming skills'', ``use technology at work (e.g., software) at least once a day'', and have an ``approval rate of at least 95\%''. These filters were chosen following the previous studies recommendations~\cite{russo2022recruiting}. 

To ensure that responses were grounded in real-world professional experience, the opening section of the survey instructed participants to describe actual use cases involving LLMs in their daily development workflows. Participation was voluntary, and all participants were compensated in accordance with Prolific’s ethical compensation guidelines. Data collection was conducted over a one-week period, resulting in a diverse sample of software professionals with varying levels of experience, roles, and technical domains.

To maintain data quality, we implemented built-in screening and validation checks throughout the survey~\cite{danilova2021you, reid2022software, alami2024you}. These included: (1) \textit{Domain knowledge validation}, where participants were asked basic programming questions (e.g., identifying common data structures or algorithm types) to confirm their familiarity with core software engineering concepts, (2) \textit{Attention checks}, where participants were explicitly instructed to enter specific words such as ``LLM'' in open-text fields to verify they were reading and understanding the questions, and, (3) \textit{Screening logic}, which required participants to confirm their role, years of experience, and familiarity with LLMs before continuing. Incomplete, inconsistent, or failed-quality-check responses were excluded from the final dataset.

\subsection{Filtering}

We obtained 297 responses. We carefully reviewed and filtered our data to include only valid responses. We removed participants who did not complete the survey (119 cases) and those who failed the prescreening questions (47 cases). We also dropped answers that failed the attention check question and checked for patterns such as selecting the same option for all multiple-choice items (0 cases). Moreover, we analyzed the time taken to complete the survey to identify lower outliers and removed one response that appeared to have been generated using an LLM instead of based on actual experience. These filtering criteria are not mutually exclusive—some participants failed more than one. After this filtering process, we end up with 131 valid responses for analysis.

\subsection{Data Analysis}

This study combined quantitative and qualitative data collected through the survey. Demographic responses were analyzed using descriptive statistics~\cite{george2018descriptive} to summarize the participant sample. Additionally, open-ended answers were explored through thematic analysis~\cite{cruzes2011recommended} to identify patterns in how software professionals use LLMs in their daily coding activities, as well as the perceived benefits and limitations of this use. Two researchers independently participated in the analysis process. To support the organization and initial coding of the qualitative data in this preliminary study, we used ChatGPT-4.0 to assist with identifying and grouping common types of usage based on participants' open-text responses.

We developed a prompt to help the model detect and classify specific coding activities where LLMs were mentioned. These activities reflected those explicitly asked about in the survey and those emergent from participant narratives. The classification was grounded in common real-world coding tasks obtained from previous studies~\cite{hou2024large, michelutti2024systematic}. 
The prompt provided to ChatGPT-4.0 is reproduced below:

\begin{promptbox}{Analysis Supporting Prompt}

\textbf{Instruction}: You are a software engineering researcher analyzing qualitative responses from professionals about their use of LLMs in daily development tasks. For each response, identify excerpts that describe concrete uses of LLMs and assign them to one or more of the following categories: Code Generation and Automation, Debugging and Troubleshooting, Learning Support, Management Activities, Testing Support, Requirements Support, or General SE Tasks. \\

\textbf{Output Requirements}:
Return the results in an Excel file with three columns:
\begin{itemize}
    \item 'Participant ID' – a unique identifier (e.g., P034)
    \item 'Use Category' – the assigned label (e.g., “Testing Support”)
    \item 'Extracted Snippet' – the exact portion of text that justifies the classification. \\
\end{itemize}

\textbf{Note}: Do not infer meaning or label vague comments. Only classify statements that describe explicit uses.
\end{promptbox}

Similar prompts were defined to extract mentions of perceived benefits, limitations, and current needs. ChatGPT-4.0 was used to flag these statements, which were then manually reviewed and finalized by the research team. In general, the analytical process was conducted in four main stages:
\begin{itemize}
    \item \textbf{Familiarization}: Researchers read responses to understand how participants used LLMs and developed the prompt to guide ChatGPT-4.0 in identifying meaningful excerpts.
    \item \textbf{LLM Tagging and Review}: ChatGPT-4.0 was used to label responses. Then, two researchers reviewed all outputs, correcting errors and removing misclassification.
    \item \textbf{Consensus Coding}: Researchers compared their analyses and resolved differences through discussion and consensus to finalize the categories.
    \item \textbf {Consolidation}: The final themes were constructed by grouping related responses under each category and identifying common trends between participant experiences.
\end{itemize}

Although this study involved responses from 131 participants, the diversity and richness of the narratives supported an in-depth analysis. Furthermore, we monitored thematic saturation during the manual review phase and found that the data reached saturation, since no novel usage categories or challenges appeared in the final round of analysis.

\subsection{Ethics}
All procedures followed institutional ethical guidelines and were approved by the third author's institution. Participation was voluntary and anonymous, with informed consent obtained during the survey. No personal data was collected. Since the dataset for this research is still being consolidated (i.e., this is a preliminary study), the full dataset is not publicly available; however, anonymized quotes are included in the results section to support transparency.

\section{Results} 
\label{sec:findings}

We begin by presenting the demographics and general profile of our study participants. The dataset comprises 131 software engineers from 14 countries, with the largest shares from the United States, South Africa, and the United Kingdom. Participants reported diverse levels of experience, most commonly within the 1–5 year (38.9\%) and 5–10 year (29.8\%) ranges. Common professional roles included Data Scientist or Machine Learning Engineer, Full-Stack Developer, and Front- or Back-End Developer. The most frequently reported software domains were Web Applications, Artificial Intelligence and Machine Learning, and Cloud and Infrastructure.

In terms of gender, 59.5\% identified as male, 39.7\% as female, and 0.8\% preferred not to disclose. Regarding the use of large language models (LLMs) for coding, 115 participants (87.8\%) had used ChatGPT, 75 (57.3\%) GitHub Copilot, 53 (40.5\%) Google Bard/Gemini, 38 (29.0\%) Claude, and 16 (12.2\%) reported using other models. Formal education or training on AI or LLMs prior to their use was reported by 62 participants (47.3\%), while 69 (52.7\%) indicated that they had no prior training. When asked about perceived effectiveness in improving coding efficiency, 53 participants (40.5\%) rated LLMs as \textit{extremely effective}, 62 (47.3\%) as \textit{very effective}, 13 (9.9\%) as \textit{moderately effective}, 3 (2.3\%) as \textit{slightly effective}, and none (0\%) as \textit{not effective}. Table~\ref{tab:Demographics} summarizes participants’ demographic and experiential characteristics.
\begin{table}[ht]
\centering
\caption{Participant Demographics and LLM Usage (N=131)}
\renewcommand{\arraystretch}{1}
\label{tab:Demographics}
\footnotesize
\begin{tabular}{llr}
\hline\noalign{\smallskip}

\multirow{3}{*}{\textbf{Gender}} 
& Male & 78 (59.5\%) \\
& Female & 52 (39.7\%) \\
& Prefer not to disclose & 1 (0.8\%) \\ \midrule

\multirow{5}{*}{\textbf{Country}} 
& United States & 47 (35.9\%) \\
& South Africa & 20 (15.3\%) \\
& United Kingdom & 19 (14.5\%) \\
& Canada & 8 (6.1\%) \\
& Other countries & 37 (28.2\%) \\ \midrule

\multirow{5}{*}{\textbf{Experience}} 
& Less than 1 year & 5 (3.8\%) \\
& 1--3 years & 35 (26.7\%) \\
& 3--5 years & 36 (27.5\%) \\
& 5--10 years & 28 (21.4\%) \\
& More than 10 years & 27 (20.6\%) \\ \midrule

\multirow{7}{*}{\textbf{Role}} 
& Data Scientist / ML Engineer & 39 (29.8\%) \\
& Full-Stack Developer & 30 (22.9\%) \\
& Front-End Developer & 16 (12.2\%) \\
& Back-End Developer & 15 (11.5\%) \\
& DevOps Engineer & 8 (6.1\%) \\
& QA Engineer & 5 (3.8\%) \\
& Other & 18 (13.7\%) \\ \midrule

\multirow{8}{*}{\textbf{\begin{tabular}[c]{@{}l@{}}Software\\ Domain\end{tabular}}} 
& Web Applications & 85 (64.9\%) \\
& AI / Machine Learning & 54 (41.2\%) \\
& Cloud / Infrastructure & 51 (38.9\%) \\
& Enterprise Software & 45 (34.4\%) \\
& Mobile Applications & 42 (32.1\%) \\
& Embedded Systems & 37 (28.2\%) \\
& Financial / Trading Systems & 24 (18.3\%) \\
& Game Development & 9 (6.9\%) \\ \midrule

\multirow{5}{*}{\textbf{LLMs Used for Coding}} 
& ChatGPT & 115 (87.8\%) \\
& GitHub Copilot & 75 (57.3\%) \\
& Google Bard / Gemini & 53 (40.5\%) \\
& Claude & 38 (29.0\%) \\
& Other & 16 (12.2\%) \\ \midrule

\multirow{2}{*}{\textbf{Formal Training on LLMs}} 
& Yes & 62 (47.3\%) \\
& No & 69 (52.7\%) \\ \midrule

\multirow{5}{*}{\textbf{Perceived Effectiveness}} 
& Extremely effective (5) & 53 (40.5\%) \\
& Very effective (4) & 62 (47.3\%) \\
& Moderately effective (3) & 13 (9.9\%) \\
& Slightly effective (2) & 3 (2.3\%) \\
& Not effective (1) & 0 (0.0\%) \\

\noalign{\smallskip}\hline
\end{tabular}
\end{table}

\subsection{Using LLM for Coding Tasks}

Software professionals reported using LLMs in several coding-related activities to support various tasks, such as writing new code, fixing bugs, learning syntax, and generating tests. Below, we present categories of usage that emerged from our preliminary analysis with supporting quotations grounded in real-world practice.

\begin{itemize}

\item \textbf{Code Generation and Automation}:
LLMs were frequently used to generate new code or automate development tasks, significantly reducing the time spent on routine or boilerplate work. Developers typically provided high-level goals or problem descriptions, and LLMs returned usable code segments, often in the form of complete functions or scripts. This support enabled practitioners to focus on higher-level design decisions or project integration rather than implementation details. For instance, P001 emphasized the usefulness of LLMs in simplifying daily development activities. 
Likewise, P002 noted the efficiency gains. 
The examples described below reflect a shift in how developers approach coding tasks, increasingly offloading initial implementation to AI.

\begin{participantbox}
\textbf{\textcolor{darkpurple}{P001}:} \textit{``read documentation, develop code to make my day-to-day easier.''}

\begin{center}
\rule{0.5\linewidth}{0.4pt}
\end{center}

\textbf{\textcolor{darkpurple}{P002}:} \textit{``[...] ChatGPT helps me create codes in Python quickly, thus saving me time.''}
\end{participantbox}

\item \textbf{Debugging and Troubleshooting}:
Participants described relying on LLMs to diagnose and resolve bugs, especially when faced with unfamiliar code or obscure errors. By pasting code snippets or describing unexpected behavior, developers could engage the LLM in pinpointing issues and proposing step-by-step fixes. This process mirrored traditional peer support, with the model acting as a knowledgeable assistant. P003 
and P012 also highlighted how the tool was used for resolving deeper integration issues. 
These interactions suggest that LLMs are becoming an integral part of developers' error-resolution workflows.

\begin{participantbox}
\textbf{\textcolor{darkpurple}{P003}:} \textit{``I used ChatGPT to assist with debugging a PHP script [...]''}

\begin{center}
\rule{0.5\linewidth}{0.4pt}
\end{center}

\textbf{\textcolor{darkpurple}{P012}:} \textit{``The last time was for debug a transactional problem with the framework I use to access my database.''}
\end{participantbox}

\item \textbf{Learning Support}:
LLMs were not only coding assistants but also acted as learning tools for many participants, particularly when exploring new languages or frameworks. Developers used LLMs to ask conceptual questions, clarify syntax, and understand code behavior, often treating the interaction as a dialogue with a tutor. The model’s ability to explain reasoning in steps and respond iteratively helped learners deepen their understanding of the material. P073 
and P110 described using prompts to guide the model toward a didactic response. 
This highlights the growing role of LLMs in personalized, on-demand learning within development environments.

\begin{participantbox}
\textbf{\textcolor{darkpurple}{P073}:} \textit{``When I am not completely familiar with the programming language, I tend to use AI to assist with coding.''}

\begin{center}
\rule{0.5\linewidth}{0.4pt}
\end{center}

\textbf{\textcolor{darkpurple}{P110}:} \textit{``Wanted an explanation of how a certain code worked. I guided the model's reasoning process by including steps in the prompt.''}
\end{participantbox}

\item \textbf{Testing Assistance}:
Several developers leveraged LLMs to support software testing tasks, particularly in generating unit tests or outlining validation strategies. Rather than replacing their judgment, LLMs provided a starting point that developers then reviewed and adapted. The tools were used to scaffold test cases aligned with function logic and to explore edge conditions or integration scenarios indicated by P005 and 
P010. 
These practices illustrate how LLMs are expanding their role into quality assurance, offering structured yet flexible input that accelerates the test-writing process while preserving developer oversight.

\begin{participantbox}
\textbf{\textcolor{darkpurple}{P005}:} \textit{``When creating unit tests for my code, I use LLM as a guide tool and as a starter.''}

\begin{center}
\rule{0.5\linewidth}{0.4pt}
\end{center}

\textbf{\textcolor{darkpurple}{P010}:} \textit{``I also use ChatGPT to write unit-tests for swift code.''}
\end{participantbox}

\end{itemize}

These findings demonstrate the varied uses of LLMs in coding workflows. Rather than relying on these tools for a single purpose, participants described using them across distinct stages of programming. Considering this integration, most of the participants (83\%) reported using generated code as a reference and modifying it, treating LLMs as starting points rather than final solutions. Around 66\% said they ask for explanations before implementing code, using LLMs to better understand logic or unfamiliar concepts. About 55\% use LLMs for brainstorming but write the code manually, engaging them as creative partners without relying on direct output. Finally, 36\% directly copy-paste generated code, often for straightforward or repetitive tasks.

\subsection{The Promise, the Pitfalls, and the Pragmatism}

In our preliminary analysis, we identified several advantages of using LLMs in coding tasks. The most frequently reported benefits included increased productivity, automated debugging, and improved code understanding. Many participants emphasized how LLMs help them work faster by reducing the time spent on repetitive or boilerplate tasks. For instance, P002 shared: 

\begin{participantbox}
\textit{``Speed to code quickly without needing to search.''}
\end{participantbox}


Participants also highlighted how LLMs assist in locating and resolving bugs more efficiently, often replacing the need to browse external forums or documentation. For example, P101 noted:

\begin{participantbox}
\textit{``I do not have to go to stack overflow anymore and search for solutions to my bug.''}
\end{participantbox}


Another commonly reported benefit was the ability to better understand complex or unfamiliar code. P109 commented: 

\begin{participantbox}
\textit{``LLMs help to understand code written by others developers.''}
\end{participantbox}


Other reported benefits of LLMs for coding included automated code documentation, efficient API and library usage, accelerated learning, reduction in cognitive load, testing and validation support, improved code refactoring, cross-language code translation, improved code quality, customization through prompt engineering, rapid prototyping, and better code completion.

In addition, we identified some limitations of using LLMs in coding tasks. The concerns most frequently reported included limited context awareness, struggles with complex logic and algorithms, and the generation of buggy or insecure code. Many participants emphasized that LLMs often fail to retain important details from previous inputs, forcing them to repeat instructions or reframe prompts. For instance, P053 shared: 

\begin{participantbox}
\textit{``Besides claude, all the other LLM's lack context and will always provide code that is prone to not work and is very generic.''}
\end{participantbox}


Participants also highlighted how LLMs struggle to reason through complex problems or implement more advanced programming logic. For example, P006 noted: 

\begin{participantbox}
\textit{``For more complex coding tasks the LLM still gives errors or wrong solutions. I have tried to solve larger problems using the LLM and have not found satisfactory solutions.''}
\end{participantbox}


Another commonly reported concern was the presence of subtle bugs or security flaws in the generated output. P104 commented:

\begin{participantbox}
\textit{``LLMs can sometimes provide code that does not work as intended or introduces new bugs.''}
\end{participantbox}


Other reported limitations of LLMs for coding included hallucination and incorrect responses, issues with code optimization, bias in code, over-reliance, reduction of critical thinking, legal and ethical concerns, and high computational costs.

While participants recognized both the advantages and limitations of LLMs in coding tasks, their responses revealed a cautious stance toward using these tools in high-stakes scenarios. When asked whether they would feel comfortable deploying LLM-generated code in production systems without human review, most participants expressed hesitation. Out of 131 valid responses, 69 participants (53\%) answered ``no'' while 28 (21\%) said they would consider it ``only for non-critical tasks''. Only 34 participants (26\%) stated they would be comfortable doing so without human oversight. These findings reinforce that, despite the practical benefits of LLMs, such as increased speed and support for debugging, developers remain concerned about reliability, correctness, and accountability, especially in production-level software.

\rqone[
    \tcblower
    \textbf{Answer:} Most developers use LLMs as supportive tools—83\% modify generated code, 66\% seek explanations, and 55\% brainstorm ideas while coding manually. Only 36\% directly copy-paste code, typically for simple tasks. Professionals view LLMs as pragmatic collaborators that can speed up and enhance various coding tasks, especially in ideation, learning, and repetitive programming. However, their use is tempered by awareness of the tools’ current limitations and the risks of over-reliance. LLMs are thus embraced not as replacements but as assistive tools—useful, but questionable, and best deployed under human oversight.
]{}

\section{Discussions} 
\label{sec:discuss}

Our findings confirm and extend existing research showing that LLMs are being actively integrated into software development workflows to support a range of tasks such as writing new code, fixing bugs, generating tests, and learning syntax~\cite{hou2024large, michelutti2024systematic, nam2024using, santos2024we}. Participants in our study described how LLMs helped them reduce time spent on routine or boilerplate tasks and accelerate comprehension when working with unfamiliar code. These uses reflect the growing perception of LLMs as general-purpose tools across the development lifecycle, as suggested in recent reviews~\cite{fan2023large, hou2024large, wang2024software}. However, our results contribute new practitioner-level detail by showing how this integration occurs across distinct stages of development, with participants not only generating or editing code, but also using LLMs to brainstorm, understand logic, and scaffold unit tests.

In contrast to earlier studies that focused on specific tools, roles, or regions~\cite{santos2024we, nam2024using}, our study draws from a broader and more diverse participant pool. With 131 professionals from 14 countries working in various roles and domains, we provide updated insights into how LLMs are being used in real-world settings. Participants described combining different strategies—such as modifying generated code, prompting for explanations, and using LLMs for ideation—to engage with the tools pragmatically. This suggests that developers are not using LLMs in a uniform way but are instead developing situated practices that reflect their confidence in the tools and the criticality of the task. Such behaviors, which are less emphasized in existing literature, highlight how usage is shaped by both task-specific demands and individual preferences.

Our results also offer a more detailed view of how trust is constructed around LLM-generated code. While earlier work has acknowledged concerns about reliability and correctness~\cite{fan2023large, zheng2025towards}, we found that developers tend to treat LLMs as useful but fallible assistants. Most participants reported modifying or verifying generated outputs, and only a minority expressed confidence in using LLM-generated code in production without human oversight. These findings echo recent concerns about the risks of hallucinations and shallow reasoning~\cite{zheng2025towards}, but they also point to an emerging sense of responsibility among practitioners, who are actively negotiating when and how to rely on these tools. The reported balance between speed and scrutiny reinforces the view that trust in LLMs is dynamic and closely tied to perceptions of accountability.

Finally, our analysis reveals a wider range of both benefits and limitations than is typically highlighted in current reviews. In addition to known advantages such as productivity, debugging support, and reduced reliance on documentation~\cite{nam2024using, hou2024large}, participants in our study reported value in areas like rapid prototyping, cross-language translation, and support for test generation and code comprehension. On the other hand, concerns extended beyond correctness, including struggles with context retention, complex logic, and subtle bugs. These findings demonstrate that while LLMs are being integrated into real-world workflows, they remain constrained by technical and cognitive limitations that require ongoing developer oversight. Our study contributes to current literature by grounding these observations in first-hand accounts of everyday development work, offering an updated understanding of how professionals use and evaluate LLMs in practice.

\subsection{Implications}
Our study has implications for both software engineering research and industry practice. As LLMs become embedded in professional development workflows, understanding how they are used, trusted, and evaluated by practitioners when working with code is important to guide future tool design, empirical investigation, and organizational adoption.

\textbf{Implications for Research.} This study advances the literature on LLM-assisted software engineering by providing empirical evidence on real-world usage, grounded in diverse practitioner experiences. Therefore, our results offer a foundation for more targeted investigations into developer–LLM interactions, particularly in areas where current understanding remains limited. For instance, future research can explore how trust in LLMs develops over time, how different prompting strategies evolve with experience, and how developers’ roles and backgrounds shape usage. Furthermore, our findings highlight the need to study the social and collaborative dimensions of LLM use, such as how teams negotiate accountability, validate AI-generated outputs, or balance efficiency in software projects. These directions would support moving beyond individual tool evaluations toward a socio-technical understanding of LLMs in software engineering.

\textbf{Implications for Practice.} For practitioners and organizations, our findings offer a perspective about the realities of LLM usage in coding workflows. Developers in our study reported using LLMs in a modular and selective way, frequently adapting, verifying, or discarding generated code rather than using it wholesale. This suggests that LLMs are most effective when treated as flexible, semi-autonomous collaborators rather than end-to-end automation tools. Organizations can use these insights to set expectations around LLM adoption, encouraging practices such as iterative prompt refinement, human-in-the-loop review, and shared strategies for mitigating risk. Moreover, our data show that trust in LLMs is closely tied to context and perceived task criticality, which highlights the importance of usage guidelines and role-specific support. Hence, organizations should consider how LLMs can complement (not replace) existing knowledge-sharing practices, documentation processes, and team-level review mechanisms.

\subsection{Threats to Validity} 
\label{sec:limitations}

As with any empirical investigation, this study presents some limitations~\cite{ralph2020empirical}. First, while the survey sample includes a diverse group of software professionals across countries, roles, and experience levels, it is not statistically representative of the global population of software developers, and we do not claim statistical generalization. Instead, our goal is to provide preliminary and analytically generalizable insights into how LLMs are used in real-world coding tasks. Second, this paper presents early findings from an ongoing study, and our results reflect emerging patterns that will be explored further in future work. Third, while we used ChatGPT-4.0 to support qualitative data organization and labeling, all outputs were manually reviewed by at least two researchers to reduce misclassification. Nevertheless, the use of an LLM for qualitative analysis introduces a threat to reliability~\cite{lecca2024applications}. Lastly, although we followed best practices for recruitment and implemented validation checks, the use of self-reported data and a crowdsourcing platform presents inherent risks of misrepresentation. We attempted to mitigate these risks through prescreening questions, attention checks, and manual data review.

\section{Conclusions} 
\label{sec:conclusions}

Our study explored how software professionals integrate LLMs into their coding workflows, based on the experience of 131 individuals. Our findings revealed that LLMs are used for a variety of coding-related tasks, including code generation and automation, debugging and troubleshooting, learning support, unit testing, and requirements clarification. Reported benefits included increased productivity, improved code quality and completion, enhanced understanding of unfamiliar code, automated debugging and documentation, efficient use of APIs and libraries, support for learning and refactoring, reduced cognitive load, cross-language translation, testing and validation assistance, customization via prompt engineering, and rapid prototyping. Despite the advantages, participants expressed concerns such as hallucinations, lack of context retention, insecure code, poor optimization, bias in outputs, over-reliance on LLMs, reduced critical thinking, legal and ethical issues, and high computational costs. While many professionals treat LLMs as helpful collaborators, they remain cautious about deploying LLM-generated code without human oversight. These early findings offer a grounded understanding of current LLM practices and highlight the importance of further research into trust, responsible usage, and the evolving role of AI in software development. In future work, we aim to expand this survey to explore how usage patterns vary across experience levels, team contexts, and software domains.



\balance

\bibliographystyle{ACM-Reference-Format}
\bibliography{bib}

\end{document}